# A Tactile Closed-Loop Device for Musical Interaction

Staas de Jong
LIACS
Media Technology program, Leiden University
staas@liacs.nl

## ABSTRACT
This paper presents a device implementing a closed tactile loop for musical interaction, based on a small freely held magnet which serves as the medium for both input and output. The component parts as well as an example of its programmable behaviour are described.

## 1. INTRODUCTION
Musical expression needs its own specialized instruments, and haptic feedback is essential to their development. Both the promise and necessity of variable force feedback for improving the expressive potential of electronic musical instruments have been noted, for instance in [6] and [1], respectively.

Another important notion concerns the fruitfulness of exploring and exploiting well-developed and readily available general input devices as musical controllers. This has been stressed in [7], illustrated by the example of a digitizing tablet. Another example is the use of force feedback joysticks and mice in [3].

At the same time, an integrated approach to musical instrument design - as for instance advocated in [5] - can motivate the conception and construction of controllers specifically intended for musical use. It is with this in mind that this paper presents a small but complete platform for investigating touch as a shared medium between performer and instrument.

## 2. THE DEVICE
### 2.1 Sensing
The device is operated by a small, handheld permanent magnet (Figure 1). The proximity of this magnet to the device surface is used as input: an LDR (light-dependent resistor) measures the gradual blocking of environment light caused by changes in its distance.

The LDR circuit's measurements are converted to digital input using a MIDI-based solution (developed at the Royal Conservatory of The Hague) giving a resolution of 7 bits at a sampling rate of approximately 100 Hz. This input is read out for further processing by a Max/MSP patch running on a suitable host computer.

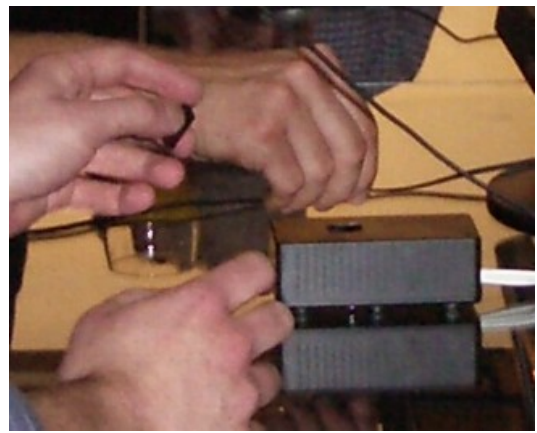

**Figure 1. The device during use.**

In order to transparently adapt to a wide range of lighting situations, only the range of values actually coming in is used: calibration is performed simply by doing a full movement towards the sensor after switching on. The resolution is further decreased by an additional detection threshold of 5%, to avoid slow fluctuations in environment light (for instance, passing clouds) being detected as an object closing in. A low-sensitivity backup input has been added for situations where the 50 Hz. flickering of light bulbs cannot be avoided.

After these and various other improvements, the sensor achieves an effective resolution of 105 steps in good lighting conditions (daylight or stable electrical light). This over a typical distance range of 0 - 4 cm. straight above the sensor; where the maximum distance can vary depending on the angle of incoming environment light.

### 2.2 Force feedback
The sensor is mounted directly onto an electromagnet, which is used to apply a variable force to the freely held permanent magnet. This force is invisible in the sense that it is transferred through air without the use of moving mechanical parts. In this, the use of electromagnetism here differs from the approach in [2].

The electromagnet was taken from an RXAB010 electro-mechanical relay and is used with a maximum current of 2 A. It is capable of generating a rejecting force sufficient to make single-handedly touching it with the permanent magnet nearly impossible; fluctuations in this force can be perceived in a range of 0 - 10 cm. above the coil. (Consequently, the playability range of the device is determined by its sensing range, which is smaller.) The electromagnet is capable of attraction as well as rejection, without the need for current reversal: when current is low, magnetization of its core by the nearby permanent magnet generates an attracting force.

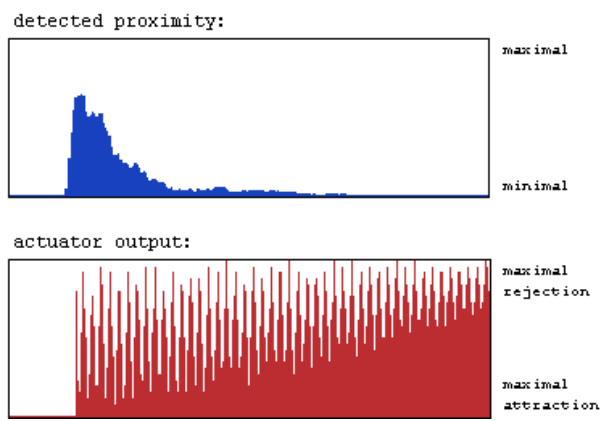

**Figure 2. An example behaviour for controlling percussive sound. (Time increases towards the right, 100 samples/sec.)**

The same MIDI device mentioned before controls the circuitry of the electromagnet. Here a fixed calibration has been used for digital output, resulting in 26 steps of resolution between maximal attraction and maximal rejection, at a sampling rate of 200 Hz. In order to be able to use Max/MSP's powerful facilities for generating complex signals, the output object for controlling the electromagnet takes an arbitrary audio-rate signal as input. There is no filtering or averaging stage before downsampling in order to allow for the generation of sharp block pulses.

## 3. EXAMPLE OF PROGRAMMABLE BEHAVIOUR

Besides handling the input and output for the sensor/actuator, the Max/MSP patch also generates sound. In one implemented example behaviour, passing a certain proximity threshold with the handheld magnet triggers synthesis of a low percussive sound. Simultaneously, a maximal rejecting force is activated, heavily modulated by a gradually decreasing triangular wave oscillating between rejection and attraction. When making a percussive movement towards the device, this creates a sense of touching a surface and causing a reverberation, which gradually dies away with the sound it produces.

Figure 2 shows the simultaneous tactile input and output of the system during such a percussive movement. The spikes in the actuator output can be seen reflected in the proximity input. (These and other example behaviours were successfully tried out by the visitors of an exhibition featuring the device.)

## 4. CONCLUSION AND FUTURE WORK

This paper has presented the implementation of a closed tactile loop for musical interaction, based on a freely held permanent magnet in combination with a fixed electromagnet. Work is under way to improve the technology underlying the device, especially the quality of its I/O. Also in progress is the implementation of more subtle tactile hints. Future work includes the integration of tactile behaviour with different mapping strategies [4] in order to achieve increasingly engaging control; as well as a more general goal of investigating more and more refined interaction techniques.

## 5. ACKNOWLEDGMENTS

Many thanks go to Lex van den Broek of the Royal Conservatory of The Hague for generously sharing his expertise in current control. Thanks also to Emiel de Jong for donating electromagnets.